\documentclass[nofootinbib,aps,twocolumn,superscriptaddress]{revtex4}
\usepackage{amsmath,amssymb}
\usepackage{citesort}
\usepackage{epsfig}
\usepackage{pstricks,pst-node,pst-text,pst-3d}

\newcommand{\beqa}{\begin{eqnarray}}
\newcommand{\eeqa}{\end{eqnarray}}
\newcommand{\beq}{\begin{equation}}
\newcommand{\eeq}{\end{equation}}
\newcommand{\bal}{\begin{align}}
\newcommand{\eal}{\end{align}}

\def\gsim{\ \rlap{\raise 3pt \hbox{$>$}}{\lower 3pt \hbox{$\sim$}}\ }
\def\lsim{\ \rlap{\raise 3pt \hbox{$<$}}{\lower 3pt \hbox{$\sim$}}\ }

\def\bl{\overline{\lambda}}
\def\b{{\cal B}}
\def\Bbar{\overline{B}^0}
\def\Abar{\overline{A}}

\def\GammaB{\overline{\Gamma}}

\begin{document}

\preprint{\vbox{
\hbox{}
\hbox{TECHNION-PH-2005-16}
\hbox{CMU-TH-05-12}
\hbox{hep-ph/0512148}
\hbox{December 2005} 
}}

\vspace*{48pt}
\title{
Weak phase $\alpha$ from $B^0\to a_1^{\pm}(1260)\pi^{\mp}$}

\def\addtech{Department of Physics,
Technion--Israel Institute of Technology,
Technion City, 32000 Haifa, Israel}
\def\addcmu{
Department of Physics, Carnegie Mellon University,
Pittsburgh, PA 15213}
\def\addIJS{J.~Stefan Institute, Jamova 39, P.O. Box 3000, 1001
Ljubljana, Slovenia}

\author{M. Gronau}\affiliation{\addtech}
\author{J. Zupan}
\affiliation{\addcmu}
\affiliation{\addIJS}

\begin{abstract} \vspace*{18pt}
Quasi two-body decays $B^0(t)\to a_1^{\pm}(1260)\pi^{\mp}$ identified by four charged pions 
determine a phase $\alpha_{\rm eff}$, which is equal  to the weak phase $\alpha$
in the limit of vanishing penguin amplitudes. Applying flavor SU(3) to these 
decays and to $B\to a_1K$ and $B \to K_1\pi$, with $K_1$ an admixture 
of $K_1(1270)$ and $K_1(1400)$, we derive expressions providing bounds 
on $\alpha-\alpha_{\rm eff}$. 
Higher precision in $\alpha$ may be achieved by an overall fit 
to a complete set of SU(3) related measurements. A method is sketched applying isospin 
symmetry to time-dependent invariant mass distributions in $B\to \pi^+\pi^-\pi^0\pi^0$.
\end{abstract}

\maketitle

\section{Introduction}
Hadronic $B$ decays from $\bar b\to \bar uu \bar d$ transitions provide the most direct information
about the weak phase $\alpha\equiv {\rm arg}(-V_{td}V^*_{tb}/V_{ud}V^*_{ub})$ governing 
the interference between $B^0-\bar B^0$ mixing and $B$ decay amplitudes in these transitions. 
The current determination of $\alpha$ from time-dependent CP asymmetries in $B^0\to \pi^+\pi^-, B\to\rho^{\pm}\pi^{\mp}$ and $B\to\rho^+\rho^-$ involves a combined error at a level of $10^\circ$~\cite{Charles:2004jd,Gronau:2005cz}.

Information on $\alpha$ can also be extracted from time-dependent decays 
$B^0(t) \to a_1^{\pm}(1260)\pi^{\mp}$~{\cite{Aleksan:1990ts,Harrison:1998yr}. 
Recently the Babar~\cite{Aubert:2005xi} and Belle~\cite{Abe:2005rf} collaborations 
have reported branching ratio measurements for these processes, where final
states were identified through four charged pions, 
\beq
\b(B^0\to a_1^{\pm}(1260)\pi^{\mp})  = \left\{ \begin{array}{c}
(40.2 \pm 3.9 \pm 3.9)\times 10^{-6}\,\mbox{\cite{Aubert:2005xi}}~,\cr
(48.6 \pm 4.1 \pm 3.9)\times 10^{-6}\,\mbox{\cite{Abe:2005rf}}~.\end{array} \right.
\eeq
These values, where charge-conjugation averaging is implied for initial and final 
states, are in agreement with a calculation based on naive factorization~\cite{Bauer:1986bm}.

The difficulty in extracting $\alpha$, common to all the above modes, is the presence of  
subleading penguin amplitudes with a different weak phase than that of the dominant tree 
amplitudes. This difficulty can be overcome by using symmetries, either 
isospin~\cite{Gronau:1990ka,Gardner:1998gz} or approximate flavor 
SU(3)~\cite{Gronau:1994rj,Dighe:1997wj}.
Applications of these symmetries to $B^0\to a_1^{\pm}\pi^{\mp}$ resemble applications
to $B\to\rho\pi$ data, where isospin symmetry in a Dalitz plot 
analysis~\cite{Snyder:1993mx,Aubert:2004iu} and flavor SU(3) for quasi two-body 
decays~\cite{Gronau:2004tm,Aubert:2003wr} have already been used to determine $\alpha$.  

An essential point in applying isospin to time-dependent decays into multibody final states 
is the existence of a final state which is common to several resonant channels having some 
overlap in phase space. 
This permits measuring relative phases between decay amplitudes for distinct resonant 
channels in $B^0$ and $\bar B^0$ decays. 
In $B\to \rho\pi$, decays to the final state $\pi^+\pi^-\pi^0$ involve interference of 
$B^0(\bar B^0)\to\rho^+\pi^-, B^0(\bar B^0)\to \rho^-\pi^+$ and $B^0(\bar B^0)\to \rho^0\pi^0$ 
in the three corners of 
the Dalitz plot \cite{Snyder:1993mx,Aubert:2004iu}. 

In contrast, one can readily show that in $B\to a_1\pi~ (a_1\to\rho\pi)$ the final state 
$\pi^+\pi^-\pi^0\pi^0$, common to $B^0(\bar B^0)\to a_1^+\pi^-, B^0(\bar B^0)\to a_1^-\pi^+$ and 
$B^0(\bar B^0)\to a_1^0\pi^0$, does not involve an overlap between the $a_1^{\pm}$ resonance
bands and the $a_1^0$ resonance band. Each of the three resonant amplitudes does interfere
with the dominantly longitudinal amplitudes~\cite{Aubert:2005nj}  for 
$B^0(\bar B^0)\to\rho^+\rho^-$. (In Ref.~\cite{Harrison:1998yr} cuts on $B^0\to a_1\pi$ were 
suggested to eliminate this interference.) Thus, in principle, a fit for 
$B^0(t)\to \pi^+\pi^-\pi^0\pi^0$ combining contributions from $B\to a_1\pi$ and $B\to \rho\rho$ 
could permit measuring relative phases between the three $B^0\to a_1\pi$ amplitudes. 
The absence of a penguin amplitude in the $\Delta I=3/2,~I=2$ linear combination of these 
amplitudes~\cite{Gronau:1990ka} would then enable an extraction of 
$\alpha$~\cite{Snyder:1993mx}. While in this respect the situation seems similar to 
$B\to\rho\pi$, one would face in this rather complex analysis 
the challenges of an additional $\pi^0$ in the final state and of an uncertainty in the $a_1$ 
resonance shape.

The purpose of this Brief Report is to propose an 
easier measurement of $\alpha$ in time-dependent 
decays $B^0(\bar B^0)\to a_1^{\pm}(1260)\pi^{\mp}$ with four charged pions 
in the final state, which is the cleanest signal channel to reconstruct. 
We will study these decays in the quasi two-body approximation within a complete 
set of SU(3) related processes. While we will follow an analogous study of 
$B^0(t)\to\rho^{\pm}\pi^{\mp}$~\cite{Gronau:2004tm},
a modification is required by the fact that $K_{1A}$, the SU(3) partner of $a_1(1260)$, is a 
mixture of two mass eigenstates, $K_1(1270)$ and $K_1(1400)$~\cite{Eidelman:2004wy}.

In Section II we define decay amplitudes and time-dependent decay rates for 
$B^0(\bar B^0)\to a_1^{\pm}\pi^{\mp}$  in terms of tree and penguin contributions, noting a 
measurable quantity $\alpha_{\rm eff}$ which equals $\alpha$ in the limit of vanishing penguin contributions. Section III derives upper bounds on $\alpha-\alpha_{\rm eff}$ in terms of branching ratios for $B\to a_1K, B\to K_1(1270)\pi$ and $B\to K_1(1400)\pi$. 
Section IV concludes, suggesting a determination of $\alpha$ using an overall parameter fit to a complete set of SU(3) related observables.
\section{Amplitudes and time-dependence}
We borrow notations and conventions from Ref.~\cite{Gronau:2004tm}. $B^0$ decay amplitudes
involve subscripts denoting the charge of the $a_1$, while $\Bbar$ amplitudes into charge 
conjugate states are denoted by $\overline{A}$ with the same subscripts,
\beqa
A_+ & \equiv & A(B^0\to a_1^+\pi^-)~,~~~ 
A_-  \equiv A(B^0\to a_1^-\pi^+)~,\nonumber\\
\overline{A}_+ & \equiv & A(\Bbar\to a_1^-\pi^+)~,~~~ 
\overline{A}_- \equiv A(\Bbar \to a_1^+\pi^-)~.
\eeqa
The four decay amplitudes can be expressed in terms of a ``tree" amplitude ($t$) and  a 
smaller ``penguin"  amplitude ($p$).
We adopt the c-convention, in which the top-quark has been integrated out in the 
$b\to d$ penguin transition and unitarity of the CKM matrix has  been used to move
a $V^*_{ub}V_{ud}$ term into the tree amplitude. We write
\beq
\begin{split}
A_\pm=&e^{i \gamma} t_{\pm}+ p_{\pm}~,\\
\overline{A}_\pm=&e^{- i\gamma} t_{\pm}+ p_{\pm}~, \label{Apm}
\end{split}
\eeq
where dependence on the weak phase $\gamma$ is displayed explicitly, while 
$t_{\pm}$ and $p_{\pm}$ contain strong phases.

Time-dependent decay rates for initially $B^0$ decaying into 
$a_1^\pm\pi^\mp$ are given by~\cite{Gronau:1989zb}
\beq\label{Gammat}
\begin{split}
\Gamma(&B^0(t) \to a_1^\pm\pi^\mp) = e^{-\Gamma t} \frac{1}{2}\left (
|A_{\pm}|^2 + |\overline{A}_{\mp}|^2\right )\\
&\Big[1 +(C \pm \Delta C)\cos\Delta mt 
- (S \pm \Delta S)\sin\Delta mt\Big]~,
\end{split}
\eeq
where 
\beq\label{CSdef}
C \pm \Delta C \equiv \frac{|A_{\pm}|^2 - |\overline{A}_{\mp}|^2}{|A_{\pm}|^2 + 
|\overline{A}_{\mp}|^2}~,
\eeq
and
\beq
S \pm \Delta S \equiv \frac{2{\rm Im}(e^{-2i\beta}\overline{A}_{\mp} A^*_{\pm})}
{|A_{\pm}|^2 + |\overline{A}_{\mp}|^2}~.
\eeq
Here $\Gamma$ and $\Delta m$ are the average $B^0$ width and the neutral $B$ mass 
difference, respectively. For initially $\Bbar$ decays, the $\cos\Delta mt$ and $\sin\Delta mt$ 
terms in~(\ref{Gammat}) have opposite signs. Thus, time-dependence in these decays 
determines the four quantities, $S\pm \Delta S, C\pm \Delta C$.

We now define two phases which coincide with $\alpha$ in the limit of vanishing penguin 
amplitudes~\cite{Charles:2004jd,Aubert:2003wr},
\beq
\alpha^{\pm}_{\rm eff} \equiv \frac{1}{2}\arg\left (e^{-2i\beta}\Abar_{\pm} A^*_{\pm}\right )~.
\eeq
Whereas these two phases cannot be measured separately, their algebraic average 
$\alpha_{\rm eff}$  is measurable~\cite{Gronau:2004tm}: 
\beq\label{alphaeff}
\begin{split}
 \alpha_{\rm eff} &\equiv \frac{1}{2}\left (\alpha^+_{\rm eff} + \alpha^-_{\rm eff} \right )
=\frac{1}{4}\Big[
\arcsin\left (\frac{S + \Delta S}{\sqrt{1- (C + \Delta C)^2}}\right )\\
&~~~~~~~~~~~~~~~~~~~~~~~~
+ \arcsin\left (\frac{S - \Delta S}{\sqrt{1- (C - \Delta C)^2}}\right )
\Big]~.
\end{split}
\eeq

The two shifts $\alpha-\alpha^{\pm}_{\rm eff} $ are expected to increase with the
magnitudes of the corresponding penguin amplitudes, $|p_{\pm}|$. 
The shifts may be expressed in terms of $|p_{\pm}|,~\gamma$ and corresponding
CP-averaged rates and CP asymmetries in $B^0\to a^{\pm}_1\pi^{\mp}$, 
\beq
\overline{\Gamma}(a_1^{\pm}\pi^{\mp})\equiv \frac{1}{2}(|A_{\pm}|^2)+|\overline{A}_{\pm}|^2)~,~~
{\cal A}^{\pm}_{\rm CP}\equiv \frac{|\overline{A}_{\pm}|^2-|A_{\pm}|^2}
{|\overline{A}_{\pm}|^2+|A_{\pm}|^2}~.
\eeq
One finds~\cite{Gronau:2004tm,Charles:1998qx},
\beq\label{p^2}
\cos 2(\alpha-\alpha^{\pm}_{\rm eff}) = \frac{1 - 2|p_{\pm}|^2\sin^2\gamma/\overline{\Gamma}(a_1^{\pm}\pi^{\mp})}
{\sqrt{1 - ({\cal A}^{\pm}_{\rm CP})^2}}~.
\eeq
\section{Bounds on $\alpha-\alpha_{\rm eff}$ from flavor SU(3)}\label{constraints}
The corrections $\alpha-\alpha^{\pm}_{\rm eff}$ caused by the penguin amplitudes $p_{\pm}$ may 
be bounded by relating the decays $B^0\to a_1^{\pm}\pi^{\mp}$ with corresponding $\Delta S=1$ 
decays, $B\to a_1 K$ and $B\to K_{1A}\pi$, where $K_{1A}$ is a nearly equal admixture of the 
$K_1(1270)$ and $K_1(1400)$ resonances~\cite{Eidelman:2004wy}. The bounds are effective
because of a relative factor $(\bl)^2,~(\bl = 0.23$) between the ratios of penguin-to-tree 
amplitudes in $\Delta S=0$ and $\Delta S=1$ processes. 
The bounds become more restrictive for small values of $|p_{\pm}|/|t_{\pm}|$. 
For instance, branching ratios for $B\to a_1 K$ and $B\to K_{1A}\pi$ which are 
not much larger than $\b(a_1^{\pm}\pi^{\mp})$ would imply generically $|p_{\pm}|/|t_{\pm}|\ll 1$
(see discussion below), similar to what has been observed in 
$B^0\to \rho^{\pm}\pi^{\mp}$~\cite{Gronau:2004tm}.

Applying flavor SU(3) to $B^0\to a^{\pm}_1\pi^{\mp}$ we will make two approximations, neglecting $\Delta S=1$ annihilation amplitudes which are formally $1/m_b$-suppressed~\cite{Bauer:2004ck}, 
and neglecting nonfactorizable SU(3) breaking 
corrections when relating $\Delta S=0$ and $\Delta S=1$ amplitudes.
Since we expect 
$|p_{\pm}/|t_{\pm}|$ to be small,  these approximations
have only a second order effect on the extracted value of $\alpha$.

We start by discussing upper bounds on $|\alpha-\alpha^-_{\rm eff}|$ following from decay rates 
for $B^{+,0}\to a_1K$. Under the above-mentioned approximation one has
\beqa\label{B+} 
A(B^+\to a^+_1K^0) &=& - (\bl)^{-1}\frac{f_K}{f_\pi}p_-~,\\
A(B^0\to a^-_1K^+) &=& \frac{f_K}{f_\pi}[- (\bl)^{-1}p_- +e^{i\gamma}\bl t_-]~,\label{B0}
\eeqa
where $\bl =  |V_{us}|/|V_{ud}| = |V_{cd}|/|V_{cs}| = 0.23$,
and $f_\pi, f_K$ are decay constants~\cite{Eidelman:2004wy}.
We define two ratios of CP-averaged rates for these processes and for $B^0\to a_1^-\pi^+$,
multiplied by $\bl^2$,
\beq\label{R}
{\cal R}^+_-  \equiv  \frac{\bl^2f_\pi^2\GammaB(a_1^+K^0)}{f_K^2\GammaB(a_1^-\pi^+)}~,~~~~
{\cal R}^0_-  \equiv  \frac{\bl^2f_\pi^2\GammaB(a_1^-K^+)}{f_K^2\GammaB(a_1^-\pi^+)}~.
\eeq
Superscripts and subscripts denote the charges of the $B$ meson and the $a_1$ meson in the denominator.
These definitions lead to bounds on $|p_-|/|t_-|$ as mentioned above, 
\beqa\label{bound-+}
\frac{\sqrt{{\cal R}^+_-}}{1+  \sqrt{{\cal R}^+_-}}  & \le &
\frac{|p_-|}{|t_-|} \le \frac{\sqrt{{\cal R}^+_-}}{1-  \sqrt{{\cal R}^+_-}}~,\\
\label{bound-0}
\frac{\sqrt{{\cal R}^0_-}-\bl^2}{1+  \sqrt{{\cal R}^0_-}}  & \le &
\frac{|p_-|}{|t_-|} \le \frac{\sqrt{{\cal R}^0_-}+\bl^2}{1-  \sqrt{{\cal R}^0_-}}~.
\eeqa

Eqs.~(\ref{p^2})-(\ref{B+}) imply immediately
\beq
\cos 2(\alpha-\alpha^-_{\rm eff}) =  \frac{1 - 2{\cal R}^+_-\sin^2\gamma}
{\sqrt{1 - ({\cal A}^-_{\rm CP})^2}}~,
\eeq
and therefore
\beq\label{bound-B+}
|\sin(\alpha-\alpha^-_{\rm eff})|  \le  \sqrt{{\cal R}^+_-}\sin\gamma~.
\eeq
The CP-averaged rate for (\ref{B0}) obeys~\cite{Gronau:2004tm,Fleischer:1997um}
$\bl^2(f_\pi/f_K)^2\GammaB(a_1^-K^+) \ge |p_-|^2 \sin^2\gamma$. Consequently,
\beq
\cos 2(\alpha-\alpha^-_{\rm eff}) \ge \frac{1- 2{\cal R}^0_-}{\sqrt{1 - ({\cal A}^-_{\rm CP})^2}}~,
\eeq
and therefore
\beq\label{bound-B0}
|\sin(\alpha-\alpha^-_{\rm eff})|  \le  \sqrt{{\cal R}^0_-}~.
\eeq

Similar considerations can be applied in order to obtain upper bounds on 
$|\sin(\alpha-\alpha^+_{\rm eff})|$ in terms of ratios of rates involving 
$K_{1A}$, the strange quark model $^3P_1$ partner of $a_1$,
\beq\label{R_A}
{\cal R}^+_{+A}  \equiv  \frac{\bl^2f^2_{a_1}\GammaB(K_{1A}^0\pi^+)}{f^2_{K_1}\GammaB(a_1^+\pi^-)}~,~~~~
{\cal R}^0_{+A} \equiv  \frac{\bl^2f^2_{a_1}\GammaB(K_{1A}^+\pi^-)}{f^2_{K_1}\GammaB(a_1^+\pi^-)}~.
\eeq
The SU(3) decompositions of the amplitudes in the numerators
are similar to \eqref{B+} and \eqref{B0}, 
\beqa\label{K1Ach}
A(B^+\to K_{1A}^0 \pi^+) &=& - (\bl)^{-1}\frac{f_{K_1}}{f_{a_1}}p_+~,\\
A(B^0\to K_{1A}^+\pi^-) &=& \frac{f_{K_1}}{f_{a_1}}[- (\bl)^{-1}p_+ +e^{i\gamma}\bl t_+]~.
\label{K1Aneutr}
\eeqa
This implies bounds on $|p_+|/|t_+|$ similar to (\ref{bound-+}) and (\ref{bound-0}) 
with ${\cal R}^{+,0}_-$ replaced by ${\cal R}^{+,0}_{+A}$. Instead of (\ref{bound-B+}) 
and (\ref{bound-B0}) one now has
\beqa\label{bound+B+}
|\sin(\alpha-\alpha^+_{\rm eff})| & \le &  \sqrt{{\cal R}^+_{+A}}\sin\gamma~,\\
|\sin(\alpha-\alpha^+_{\rm eff})| & \le & \sqrt{{\cal R}^0_{+A}}~.\label{bound+B0}
\eeqa

We now discuss upper bounds on ${\cal R}^+_{+A}$ and ${\cal R}^0_{+A}$ in terms of 
physical processes involving the mass eigenstates $K_1(1270)$ and $K_1(1400)$.
The state $K_{1A}$ is an almost equal admixture of these states,
\beq\label{K1A}
K_{1A} = \cos\theta K_1(1400) + \sin\theta K_1(1270)~,
\eeq
with $33^{\circ} \le \theta \le 57^{\circ}$~\cite{Suzuki:1993yc},
while the orthogonal state,
the strange SU(3) $^1P_1$ partner of $b_1(1236)$, is 
\beq\label{K1B}
K_{1B} = -\sin\theta K_1(1400) + \cos\theta K_1(1270)~,
\eeq
This implies 
\beqa
A(B^+\to K^0_{1A}\pi^+) = ~~~~~~~~~~~~~~~~~~~~~~~~~~~~~~~~\\
\cos\theta\,A(K_1^0(1400)\pi^+) + 
\sin\theta\,A(K_1^0(1270)\pi^+)~, \nonumber
\eeqa
where the two pure penguin amplitudes, identified by their final 
states, involve an arbitrary relative strong phase. Thus, one has an upper bound on $\GammaB(K_{1A}^0\pi^+)$,
\beqa\label{cosGamma}
\GammaB(K_{1A}^0\pi^+) \le  ~~~~~~~~~~~~~~
~~~~~~~~~~~~~~~~~~~~~~~~~~~\\
\Big [\cos\theta\sqrt{\GammaB(K_1^0(1400)\pi^+)}
+ \sin\theta\sqrt{\GammaB(K_1^0(1270)\pi^+)}\Big ]^2~,\nonumber
\eeqa
which determines an upper bound on ${\cal R}^+_{+A}$ defined in (\ref{R_A}).

A similar expression holds for an upper bound  on ${\cal R}^0_{+A}$,
in terms of the mixing angle $\theta$ and CP-averaged decay rates for 
$B^0\to K_1^+(1400)\pi^-$ and $B^0\to K_1^+(1270)\pi^-$. This bound can be shown to 
hold in spite of the fact that these processes involve both penguin and tree amplitudes.

Finally, one combines the two separate upper bounds on $|\alpha - \alpha^-_{\rm eff}|$ and
$|\alpha - \alpha^+_{\rm eff}|$ to obtain a bound on $|\alpha-\alpha_{\rm eff}|$,
\beq\label{cons-bound}
|\alpha - \alpha_{\rm eff}| \le \frac{1}{2}(|\alpha - \alpha^+_{\rm eff}| + |\alpha - \alpha^-_{\rm eff}|)~.
\eeq
\section{Conclusion}
We have studied the extraction of $\alpha$ from time-dependent decays 
$B(t)\to a_1^{\pm}(1260)\pi^{\mp}$
in the quasi two-body approximation.
The four observables, $S\pm \Delta S$ and $C\pm \Delta C$, determine the angle 
$\alpha_{\rm eff}$ in Eq.~(\ref{alphaeff}) up to a fourfold discrete ambiguity. 
A twofold ambiguity in $\alpha_{\rm eff}$ may be resolved either by other constraints on 
$\alpha$, or by assuming that the two added angles on the right-hand side 
of Eq.~(\ref{alphaeff}) differ by much less than $180^\circ$. This follows from 
$|{\rm arg}(t_-/t_+)| \ll 90^\circ$, valid to leading order in $1/m_b$~\cite{Bauer:2004ck}, and an assumption of small $|p_{\pm}|/|t_{\pm}|$, testable through relations such as
Eqs.~(\ref{bound-+}) and (\ref{bound-0}). 

We have used flavor SU(3) to obtain upper bounds (\ref{bound-B+}), (\ref{bound-B0}), (\ref{bound+B+}) and (\ref{bound+B0}) on  $|\alpha-\alpha^{\pm}_{\rm eff}|$. 
This requires measuring CP-averaged rates for either $B^+\to a^+_1K^0$ or $B^0\to a^-_1K^+$ 
and CP-averaged rates for
either $B^+\to K_1^0(1270)\pi^+$ and  $B^+\to K_1^0(1400)\pi^+$ or $B^0\to K_1^+(1270)\pi^-$ 
and $B^0\to K_1^+(1400)\pi^-$. 
The resulting upper bound, Eq.~(\ref{cons-bound}), assumes an
unknown relative sign between $\alpha-\alpha^-_{\rm eff}$ and 
$\alpha-\alpha^+_{\rm eff}$. 
In $B^0\to \rho^{\pm}\pi^{\mp}$ these two shifts are expected to have opposite signs 
because $|p_{\pm}|/|t_{\pm}|$ are small and ${\rm arg}(p_{\pm}/t_{\pm})$ lie in opposite hemispheres~\cite{Gronau:2004sj}, as shown in a global SU(3) fit 
to $B\to VP$ decays~\cite{Chiang:2003pm} and in QCD factorization including 
$1/m_b$-suppressed terms~\cite{Beneke:2003zv}. This reduces the bound on $|\alpha-\alpha_{\rm eff}|$ in $B^0\to \rho^{\pm}\pi^{\mp}$ by a factor of two~\cite{Gronau:2004sj}. It is unclear whether a similar argument holds in $B^0\to a_1^{\pm}\pi^{\mp}$. 

Instead of using SU(3) to obtain upper bounds on $|\alpha-\alpha_{\rm eff}|$ one may perform 
a fit to all the observables in $B^0(t)\to a_1^\pm\pi^\mp$ and in SU(3) related modes. 
This study is expected to 
reduce errors and to resolve ambiguities in $\alpha$, as has been shown in the case of $B\to \rho^\pm\pi^\mp$ by performing a $\chi^2$ fit~\cite{Gronau:2004tm}.
Since the states $K_{1A}$ and $K_{1B}$ in (\ref{K1A}) and (\ref{K1B}) mix, a complete set of
processes includes also the decay
$B^0\to b_1^{+}(1235)\pi^{-}$ described by amplitudes $t^b_{+}$ and $p^b_{+}$ in
analogy with \eqref{Apm}. Information from $B^0\to b_1^{-}(1235)\pi^{+}$ is not needed since
the corresponding $\Delta S=1$ decays $B\to b_1 K$ are unrelated to $B\to a_1 K$.

Amplitude decompositions are given in Eqs.~\eqref{Apm}, \eqref{B+}, \eqref{B0}, \eqref{K1Ach}, \eqref{K1Aneutr} and by corresponding expressions for $A(B^0\to b^+_1\pi^-),~A(B\to K_{1B}\pi)$, 
with the replacements $t_+\to t_+^b,~p_+\to p_+^b,~K_{1A}\to K_{1B}$. The total number of 
observables is seventeen, including $S\pm \Delta S,~C\pm \Delta C$ and the two CP-averaged 
rates in $B^0(t)\to a^{\pm}_1\pi^{\mp}$, the CP-averaged rates and asymmetries in 
$B^0\to b_1^+\pi^-,~a_1^-K^+,~K_1^+(1400)\pi^-,~K_1^+(1270)\pi^-$, and the rates for
$B^+\to a_1^+ K^0,~K_1^0(1400)\pi^+,~K_1^0(1270)\pi^+$.
The seventeen observables are described in terms of twelve parameters, the magnitudes and relative phases of $t_\pm,p_\pm, t_+^b,p_+^b$ and the weak phase $\alpha$. A simplification, 
$t^+_b\simeq 0$, occurs  by assuming factorization of tree amplitudes, which holds at leading order 
in $1/m_b$~\cite{Bauer:2004ck,Beneke:2003zv}, and by using the $G$-parity of $b_1$~\cite{Laplace:2001qe}. 
This implies a vanishing asymmetry in $B^0\to b_1^+\pi^-$ and a small rate for this 
process unless $p_+^b$ is enhanced by nonperturbative effects. 

We thank Fernando Palombo and  Jim Smith for helpful discussions and for motivating
this study.  This work was supported in part by the United States Department of Energy 
under Grants No.\ DOE-ER-40682-143 and DEAC02-6CH03000,
by the Israel Science Foundation under Grant No. 1052/04, and by the German--Israeli 
Foundation under Grant No. I-781-55.14/2003.

\def \ajp#1#2#3{Am.\ J. Phys.\ {\bf#1}, #2 (#3)}
\def \apny#1#2#3{Ann.\ Phys.\ (N.Y.) {\bf#1}, #2 (#3)}
\def \app#1#2#3{Acta Phys.\ Polonica {\bf#1}, #2 (#3)}
\def \arnps#1#2#3{Ann.\ Rev.\ Nucl.\ Part.\ Sci.\ {\bf#1}, #2 (#3)}
\def \art{and references therein}
\def \cmts#1#2#3{Comments on Nucl.\ Part.\ Phys.\ {\bf#1}, #2 (#3)}
\def \cn{Collaboration}
\def \cp89{{\it CP Violation,} edited by C. Jarlskog (World Scientific,
Singapore, 1989)}
\def \econf#1#2#3{Electronic Conference Proceedings {\bf#1}, #2 (#3)}
\def \efi{Enrico Fermi Institute Report No.}
\def \epjc#1#2#3{Eur.\ Phys.\ J.\ C {\bf#1}, #2 (#3)}
\def \ib{{\it ibid.}~}
\def \ibj#1#2#3{~{\bf#1} (#3) #2}
\def \ijmpa#1#2#3{Int.\ J.\ Mod.\ Phys.\ A {\bf#1}, #2 (#3)}
\def \ite{{\it et al.}}
\def \jhep#1#2#3{JHEP {\bf#1}, #2 (#3)}
\def \jpb#1#2#3{J.\ Phys.\ B {\bf#1}, #2 (#3)}
\def \kdvs#1#2#3{{Kong.\ Danske Vid.\ Selsk., Matt-fys.\ Medd.} {\bf #1}, No.\
#2 (#3)}
\def \mpla#1#2#3{Mod.\ Phys.\ Lett.\ A {\bf#1}, #2 (#3)}
\def \nat#1#2#3{Nature {\bf#1}, #2 (#3)}
\def \nc#1#2#3{Nuovo Cim.\ {\bf#1}, #2 (#3)}
\def \nima#1#2#3{Nucl.\ Instr.\ Meth.\ A {\bf#1}, #2 (#3)}
\def \npb#1#2#3{Nucl.\ Phys.\ B~{\bf#1},  #2 (#3)}
\def \npps#1#2#3{Nucl.\ Phys.\ Proc.\ Suppl.\ {\bf#1}, #2 (#3)}
\def \PDG{Particle Data Group, D. E. Groom \ite, \epjc{15}{1}{2000}}
\def \pisma#1#2#3#4{Pis'ma Zh.\ Eksp.\ Teor.\ Fiz.\ {\bf#1}, #2 (#3) [JETP
Lett.\ {\bf#1}, #4 (#3)]}
\def \pl#1#2#3{Phys.\ Lett.\ {\bf#1}, #2 (#3)}
\def \pla#1#2#3{Phys.\ Lett.\ A {\bf#1}, #2 (#3)}
\def \plb#1#2#3{Phys.\ Lett.\ B {\bf#1}, #2 (#3)} 
\def \prd#1#2#3{Phys.\ Rev.\ D\ {\bf#1}, #2 (#3)}
\def \prl#1#2#3{Phys.\ Rev.\ Lett.\ {\bf#1}, #2 (#3)} 
\def \prp#1#2#3{Phys.\ Rep.\ {\bf#1}, #2 (#3)}
\def \ptp#1#2#3{Prog.\ Theor.\ Phys.\ {\bf#1}, #2 (#3)}
\def \rmp#1#2#3{Rev.\ Mod.\ Phys.\ {\bf#1}, #2 (#3)}
\def \rp#1{~~~~~\ldots\ldots{\rm rp~}{#1}~~~~~}
\def \yaf#1#2#3#4{Yad.\ Fiz.\ {\bf#1}, #2 (#3) [Sov.\ J.\ Nucl.\ Phys.\
{\bf #1}, #4 (#3)]}
\def \zhetf#1#2#3#4#5#6{Zh.\ Eksp.\ Teor.\ Fiz.\ {\bf #1}, #2 (#3) [Sov.\
Phys.\ - JETP {\bf #4}, #5 (#6)]}
\def \zp#1#2#3{Zeit.\ Phys.\  {\bf#1}, #2 (#3)}
\def \zpc#1#2#3{Zeit.\ Phys.\ C {\bf#1}, #2 (#3)}
\def \zpd#1#2#3{Zeit.\ Phys.\ D {\bf#1}, #2 (#3)}


\end{document}